# Wetting Interactions Between Porous Carbon Hosts and Liquid Sodium-Potassium Alloys Toward Their Use in Negative Electrodes of Alkali-Metal Batteries

*Johannes Baller[1], André Hilger[2], Naiyu Qi[1], Chiara Morini[1], Andrea Cornelio[1,3], Arndt Remhof[1,4,5], Markus Osenberg[2], Ingo Manke[2], Julian Moosmann[6], Felix Beckmann[6], Gustav Graeber[1,3,*]*

*gustav.graeber@hu-berlin.de

[1]Department of Chemistry, Humboldt-Universität zu Berlin, 12489 Berlin, Germany

[2]Helmholtz-Zentrum Berlin für Materialien und Energie, Lise-Meitner-Campus, Hahn-Meitner-Platz 1, 14109 Berlin, Germany

[3]Department 3: Containment Systems for Dangerous Goods; Energy Storage, Federal Institute for Materials Research and Testing (BAM), 12205 Berlin, Germany

[4]Empa, Swiss Federal Laboratories for Materials Science and Technology, 8600 Dübendorf, Switzerland

[5]Institut für Anorganische und Analytische Chemie, Albert-Ludwigs-Universität Freiburg, 79104 Freiburg, Germany

[6]Institute of Materials Physics, Helmholtz-Zentrum Hereon, Max-Planck-Straße 1, 21502 Geesthacht, Germany

Funding: This study has received funding from the Bundesministerium für Forschung, Technologie und Raumfahrt under the grant No. 03XP0516. Additional funds were received from Fonds der Chemischen Industrie im Verband der Chemischen Industrie e. V. under the grant No. 661737. This research was supported by the Federal Federal Ministry of Research, Technology and Space (BMFTR/BMBF) as part of the FORBATT project "TomoFestBattLab", grant number 03XP0462, and by the BMFTR/BMBF project "NATTER" grant number 03XP0525C.

Abstract

Batteries with liquid alkali-metal negative electrodes offer a route to compact, high-performance energy storage. Innovation in alkali-metal management, i.e., controlled storage, release and transport of liquid alkali metal, can enable simpler and cheaper cell designs. Porous carbons have emerged as potential host materials for liquid alkali metals. Here, we study the wetting interactions between porous carbon hosts and liquid sodium-potassium alloy (NaK) as a function of carbon host morphology and surface functionalization via X-ray computed tomography. While as-received carbon samples show no affinity towards NaK, heat-treated carbon is spontaneously infiltrated with NaK filling almost the entire pore volume. We explore how forced wetting partially fills pores of NaK-repellent hosts, showing large differences in pore filling based on the average pore size of the host material. In electrochemical discharge experiments, we show that both as-received and heat-treated carbon felt enable high areal capacities beyond 40 mAh cm$^{-2}$. However, the heat-treated carbon shows ten times lower overpotential. Finally, we demonstrate how heat-treated carbon felt can enable capillary transport of NaK. In summary, this study elucidates important aspects of the interactions between liquid alkali metals and porous carbon hosts, generating insights into possible applications in liquid alkali-metal batteries.

# 1. Introduction

Batteries with alkali-metal negative electrodes promise higher energy densities compared to batteries that rely on intercalation type negative electrodes. Li-ion technology relies on such intercalation negative electrode materials. In batteries with liquid electrolytes, alkali-metal negative electrodes are challenging to implement due to issues such as continuous electrolyte degradation and dendrite formation [1]. Many solid electrolytes (SE) can better compensate for these issues due to their superior chemical stability and mechanical strength [2], [3]. Implementing alkali-metal negative electrodes and the associated gains in energy density could also make the use of sodium and potassium as active materials more viable in terms of the energy density at the cell level. While sodium and potassium can only offer a lower theoretical energy density compared to lithium, they are substantially more abundant, which can reduce global dependencies and supply chain risks [4]. While SEs help to enable alkali-metal negative electrodes, their rigidity also introduces restrictions. A good contact between the negative electrode material and the SE cannot be ensured by the electrolyte itself. Therefore, it is common that solid-state batteries rely on a high stack pressure up to hundreds of MPa that forces the alkali metal to stay in contact with the SE [5].

To address the challenges of maintaining contact between alkali metals and solid electrolytes, some commercial systems employ liquid alkali-metal electrodes, as in high-temperature sodium batteries such as sodium-sulfur (NaS) and sodium-nickel-chloride (ZEBRA) cells. Here, the sodium-metal negative electrode is liquid and is paired with a sodium-β''-alumina solid electrolyte (Na-BASE) [6]. Even though the melting point of sodium is 98 °C, these high-temperature sodium batteries usually need to be operated at 200 °C or more, with some recent research focusing on lowering this temperature to 150 °C or lower [7], [8], [9]. One advantage of liquid alkali-metal negative electrodes is that dendrite formation is less likely to occur in a liquid electrode compared to a solid metal electrode [10]. Compared to a solid alkali-metal negative electrode, a liquid alkali-metal electrode can significantly enhance the contact between the metal and the SE, without the need of elevated stack pressures. However, managing liquid alkali-metals can be challenging in terms of liquid containment, wetting interactions in the cell and the corrosiveness of liquid alkali-metals. Commercialized high-temperature sodium batteries rely on a tubular cell design and gravity to manage the content of the negative electrode. Metal shims and carbon coatings can be employed to increase the area on the SE wetted by the liquid sodium [11].

A planar cell design can simplify the complex manufacturing processes currently employed in high-temperature sodium batteries [12], [13]. Such a design could, however, not necessarily rely on gravity anymore for the liquid alkali-metal management. Especially when aiming at high areal capacities, alkali-metal management therefore becomes even more important. Porous carbon materials are an attractive option to achieve good alkali-metal management. They are low-cost and lightweight and are commercially available in multiple forms of high porosity, which have seen widespread use in other battery applications [14], [15], [16]. These porous carbon materials are intrinsically phobic against liquid alkali metals. Different wetting strategies have been shown to enable the filling of porous carbon materials, such as vacuum infiltration [14] utilization of $KO_2$ stemming from deliberate oxygen impurities to increase alkali metal to carbon adhesion [10] or chemical or heat treatments of the carbon host [17], [18]. From these options, heat treatment has shown to be a very effective and simple tool to drastically increase the wettability of carbon materials by liquid alkali-metals. In this study, we are investigating the property changes occurring due to a heat treatment of porous carbon materials and possible applications. We are studying different porous carbon hosts, using liquid sodium-potassium alloy (NaK) as a room temperature alternative to molten sodium [19].

## 2. Results and Discussion

### 2.1 Thermal Treatment of Porous Carbon Hosts

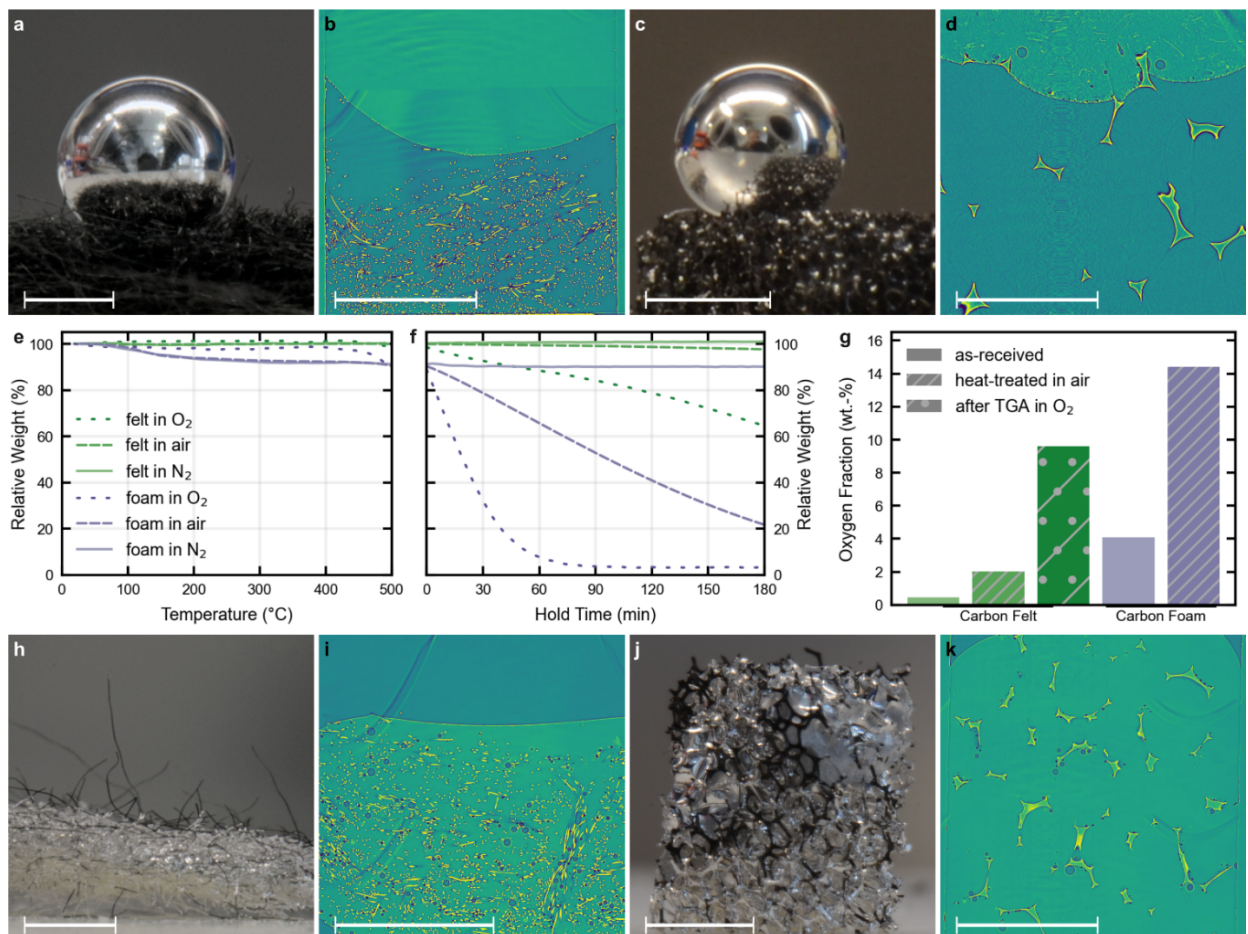

**Figure 1. Wetting interactions between liquid alkali metals and porous carbon hosts as a function of texture and thermal pretreatment. a-b**, Photograph and X-ray computed tomography (XCT) scan of a sodium-potassium (NaK) droplet resting on as-received carbon felt in a protective argon atmosphere, illustrating the strongly repelling nature of the felt against NaK. **c-d**, Photograph and XCT scan of an NaK droplet resting on as-received carbon foam showing similar repellency. **e**, Thermogravimetric analysis (TGA) of carbon felt and carbon foam with a temperature increase up to 500 °C and, **f**, a subsequent 3 h isothermal treatment. **g**, Elemental analysis obtained via energy-dispersive X-ray spectroscopy (EDX) showing that the measured atomic concentration of oxygen in the samples increases with the heat treatment. **h-i**, Photograph and XCT scan of a heat-treated carbon felt, which got spontaneously wetted by NaK, illustrating the strongly attracting nature of the heat-treated felt towards NaK. **j-k**, Photograph and XCT scan of heat-treated carbon foam after getting in contact with NaK. The NaK coats parts of the carbon structure and fills some of the pores in the photograph. In the XCT scan, NaK completely fills the pore volume except for some trapped gas bubbles. All scalebars are 2 mm. In all XCT scans, green areas are NaK, blue areas are gas, and yellow outlined areas with sharp geometry are the carbon materials.

In **Figure 1**, we show carbon felts and carbon foams, two different porous carbon hosts with two distinct morphologies, as they interact with liquid eutectic sodium-potassium alloy. When referring to specific alloy compositions we use the notation Na23K77 to indicate an alloy composition of 23 wt.-% Na and 77 wt.-% K, which is the eutectic composition. When placing a microliter-sized amount of NaK alloy onto an as-received carbon felt in an argon-filled glove box, it does not wet the felt. Instead, it beads up to a droplet and rests on top of the carbon fibers (see **Figure 1a**). This repellency is also confirmed via X-ray computed tomography (XCT) measurements (see Methods for details). In **Figure 1b**, we show a respective vertical XCT slice through an NaK droplet resting on an as-received carbon felt. The same observation of NaK repellency can be made for as-received carbon foam (**Figure 1c-d**). Both, the as-received carbon felt and the as-received carbon foam are not significantly wetted by the liquid alkali metal, despite having rather distinct morphologies and pore sizes.

In a recent study, Zhao et al. [17] treated a carbon fiber cloth in a furnace under air. The authors observed a significantly improved wettability of their fiber cloth in contact with liquid alkali metal. Along their insights, we performed dedicated thermogravimetric analysis (TGA) experiments on our porous carbon materials to also affect their wetting behavior. These TGA experiments include a temperature ramp up to 500 °C as well as a holding time of three hours. **Figure 1e-f** shows the evolution of the relative weight in percent of the carbon samples during the TGA experiments, as a function of the selected gaseous atmosphere during the heat treatment. When heating the carbon materials in a nitrogen atmosphere, both samples are stable as represented by a constant relative weight during temperature ramp up and during the subsequent hold time. In contrast, when heating the samples in an environment of air or pure oxygen, they lose weight. For the carbon felt, the mass loss at the end of the heat treatment is moderate when heated in air, reaching 2.4%. When heated in pure oxygen, the carbon felt loses 36% of its mass by the end of the treatment. For the carbon foam, the mass loss is more substantial reaching 78.3% in air, where the foam also shows deformation and shrinkage after the heat treatment. Remarkably, when heat-treating the carbon foam in a pure oxygen atmosphere, the sample burns up completely (100% mass loss) after only 90 minutes of exposure. A difference in combustion rate of the two distinct carbon samples might occur as a result of differences in their crystallinity and the quantity of defects. Research on glassy carbon has shown that a higher crystallinity and fewer defects increase the resistance of glassy carbon towards burnoff at 450 °C [20]. Raman spectroscopy measurements (**Figure S1**) show that after the heat treatment the ratio between the carbons D and G band relative intensity is lower, which has previously been shown to be an indicator of more carbon atoms in a graphitic environment [21].

To quantify possible compositional changes during the heat-treatment, we investigated various carbon samples, namely as-received, heat-treated in air, and after TGA in $O_2$, via scanning electron microscopy and energy dispersive X-ray spectroscopy (EDX). Results of this analysis are shown in **Figure 1g**. The carbon felt samples generally show a smaller oxygen weight fraction than carbon foam. Only when examining the carbon felt that previously underwent TGA in $O_2$ its oxygen weight fraction increases above the as-received carbon foam. There is a considerable increase in oxygen weight fraction after the heat treatment in air and an even stronger increase under the same heating protocol in pure oxygen for the carbon felt. The increase in oxygen content is most likely due to carbon-carbon bonds breaking and new oxygen rich end groups being introduced [17]. The initial higher oxygen content of the carbon foam is a possible reason for its greater reduction in weight due to burnoff during the heat treatment. Accompanying X-ray diffractometry (XRD) measurements (**Figure S2**) show that all samples exhibit very broad peaks that can be common in mixed amorphous/crystalline carbons with the peak position linked to a graphitic fraction of the material [22], [23].

After the heat-treatment, the wettability of carbon foam and carbon felt samples by NaK is greatly improved. (**Figure 1h-k**). This correlates with the higher concentration of oxygen-rich end groups that have been shown in the literature to react more favorably from a thermodynamic standpoint with the liquid alkali metal [17]. Due to this improved wettability, the NaK spontaneously infiltrates into the porous carbon structure. In the XCT slices shown in **Figure 1i,k**, it can be seen that the Na23K77 fills most of the open volume in the carbon, only trapping a small amount of gas. This infiltration process occurs immediately after bringing the heat-treated carbon host in contact with the NaK. The infiltration is facilitated by capillary action, which benefits from the good wettability of the carbon material towards NaK combined with the small pore size of the carbon hosts.

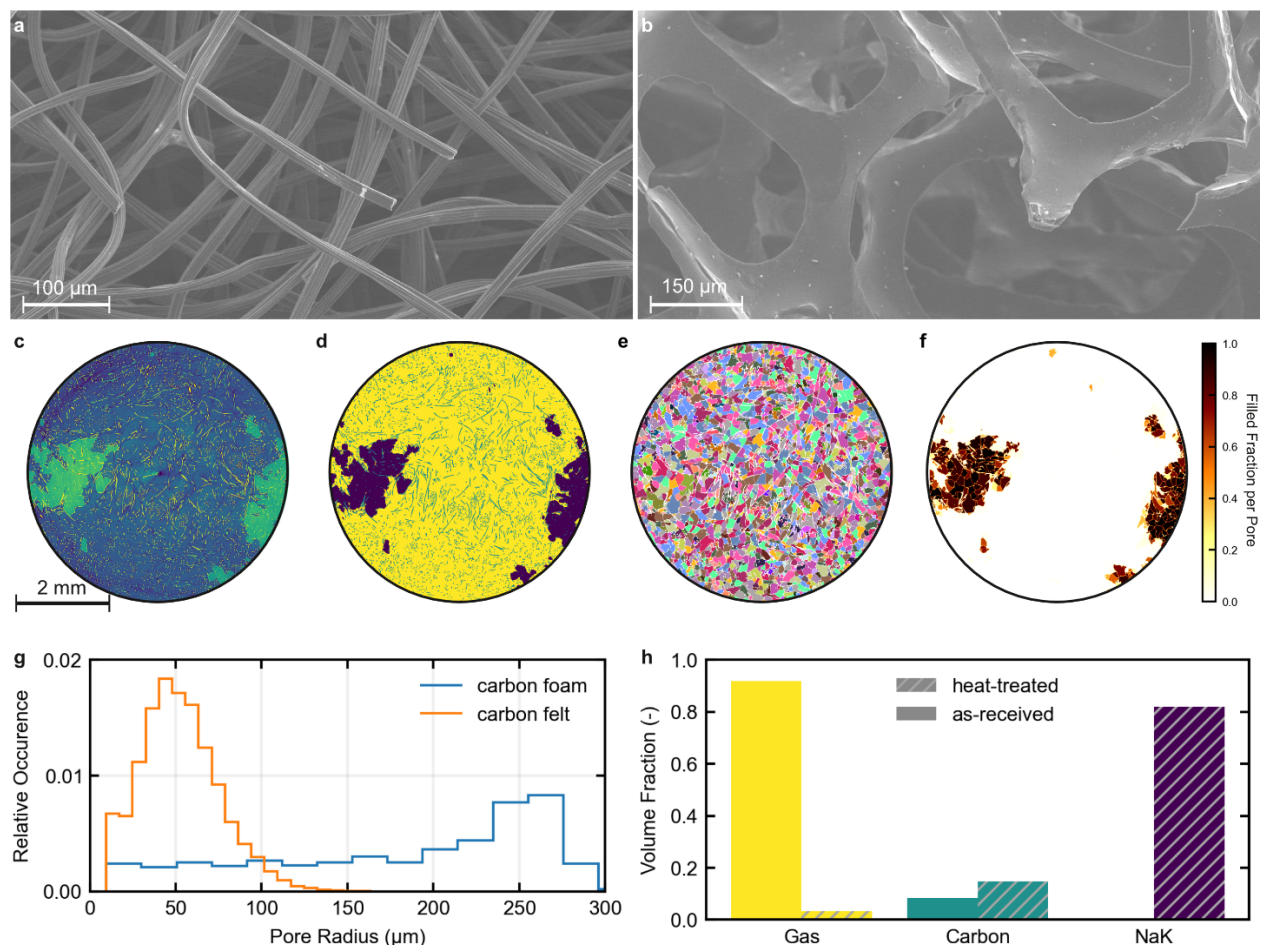

**Figure 2. Quantitative analysis of the alkali-metal wetting phenomena in porous carbon hosts via XCT**. **a**, SEM image of the carbon felt showing the micrometer-sized irregular pores being formed between the individual carbon fibers. **b**, SEM image of the carbon foam showing pores with substantially larger pore radii. **c-g**, The four steps of the XCT workflow exemplified in one horizontal slice through a carbon felt partially infiltrated with liquid NaK alloy: **c**, First step, reconstructed XCT image. **d**, Second step, segmentation of the different materials present: gas (color: yellow), carbon (teal), and liquid NaK alloy (purple). **e**, Third step, pore identification through a 3D distance transform watershed algorithm. **f**, Heatmap of degree of pore filling by NaK assessed for individual pores. **g**, Quantification of pore radii in a carbon felt (orange line) and the carbon foam (blue line), respectively. **h**, Quantification of the total volume occupied by either of the three materials present in a sample volume within an as-received carbon felt (plain bars) and a heat-treated carbon felt (hatched bars), both in contact with liquid NaK alloy.

## 2.2 Quantitative Analysis of 3D Tomography Data

XCT imaging allows to look into the bulk volume of a sample, in our case to examine the wetting behavior of liquid NaK alloy in porous carbon felts (**Figure 2a**) and carbon foams (**Figure 2b**). Reconstructed XCT data provides a three-dimensional map of the mass attenuation coefficient in the analyzed volume. The mass attenuation coefficient is linked (among other effects) to the atomic number and the density of a given material. This enables an analysis of the material distribution in the imaged volume, although imaging artefacts can complicate this relationship. Due to these measurement artifacts, a simple thresholding approach is not sufficient to reliably quantify different materials when densities are relatively similar such as with carbon and Na23K77. Instead, to enable quantitative analysis, the grayscale tomography data is processed using the supervised machine learning based segmentation tool ilastik [24]. In our case, this results in a volume with three different integer values corresponding to voxels that are categorized as gas, carbon or NaK. In the datasets used in this study, this approach yields a classification more consistent with a visual assessment than a thresholding approach.

This process is detailed in **Figure 2c-f.** Each panel shows the same horizontal slice through the same volume throughout different XCT processing steps. **Figure 2c** shows the original reconstructed tomography data. **Figure 2d** shows the segmented data where yellow is gas, teal is carbon and purple is NaK. **Figure 2e** and **f** show examples of quantification steps. To evaluate pore sizes, a 3D distance-transform watershed analysis is performed using the MorphoLibJ Fiji plugin [25]. This results in each voxel being assigned to a thus identified pore. The detected pores are shown in **Figure 2e**. Finally, **Figure 2f** shows to what degree each of the identified pores is filled with NaK. The calculation considers the voxels marked as belonging to a given pore in the 3D volume (**Figure 2e**) and divides the number of voxels detected to be NaK (purple in **Figure 2d**) by the total number of voxels. Areas that are green in **Figure 2c** or purple in **Figure 2d**, i.e., contain NaK, show mostly filled pores in this analysis. Other parts of the volume show next to no pore filling.

This methodology allows for a range of different analyses. The pore size distribution of as-received carbon felt and as-received carbon foam is shown in **Figure 2g**. This analysis considers the total number of voxels detected by the distance transform watershed analysis exemplified in **Figure 2e** and arrives at a pore diameter by assuming a spherical pore geometry. The carbon foam exhibits a significantly larger average pore size and wider pore size distribution. **Figure 2h** details the material distribution in a control volume in the carbon samples shown in **Figure 1b** and **i**. The as-received sample shows no NaK in the bulk volume. In contrast, almost all of the prior gas volume is filled with NaK in the heat-treated sample within the resolution limit of the XCT and the segmentation approach. Differences in the carbon volume fraction might arise from capillary forces from the liquid onto the felt, differences in the recognition of the boundary between two materials during the segmentation, and the manual sample preparation. These methods allow us to differentiate between pore filling by NaK or gas and to quantify wetting interactions in different porous host materials via a non-destructive approach.

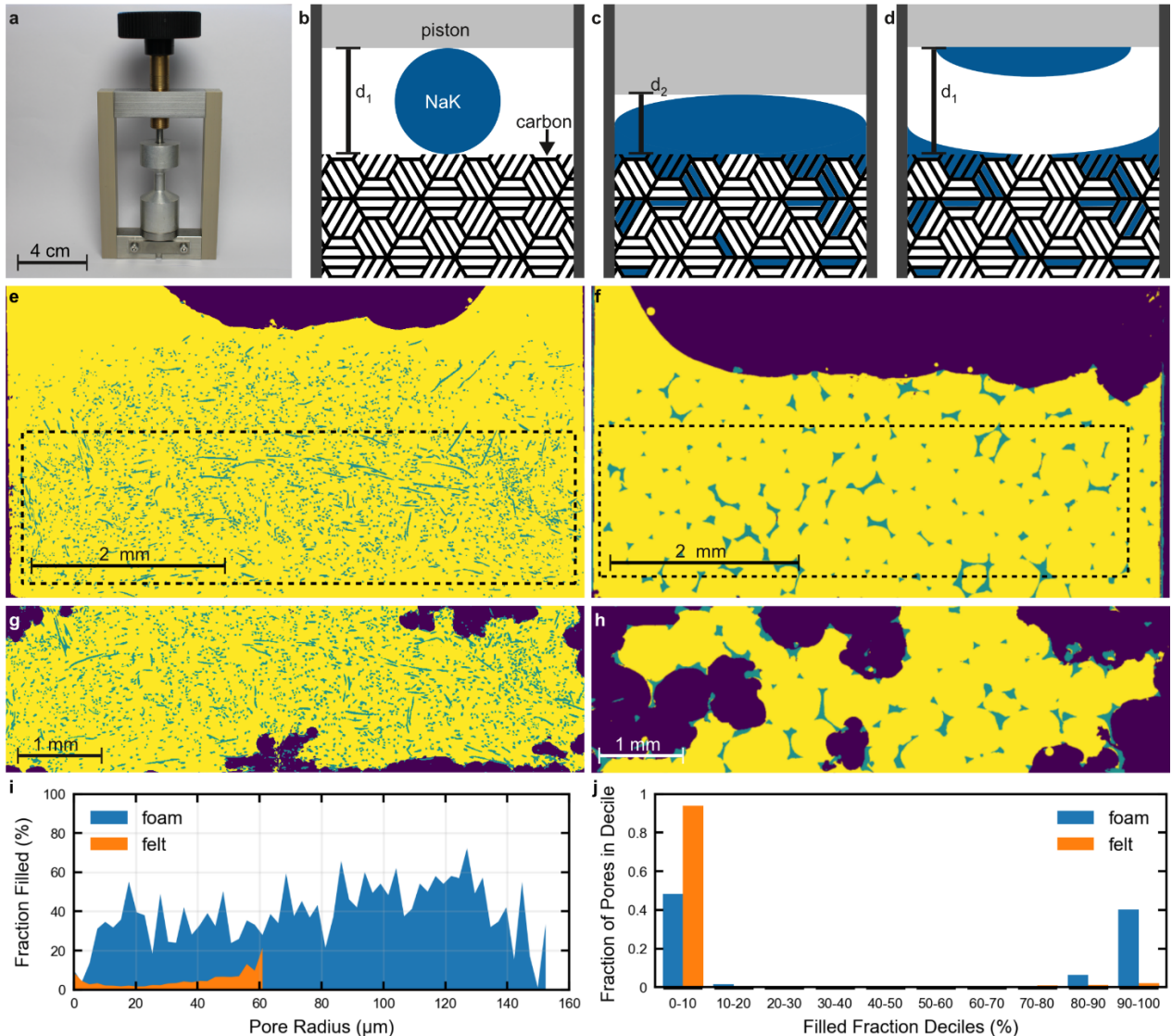

**Figure 3. Forced wetting interactions between as-received, alkali-metal repellent carbon hosts and liquid NaK droplets**. **a**, Photograph of the tomography cell with an adjustable piston. **b-d**, Schematic drawings of the interactions between the NaK droplet with the porous hosts showing, **b**, the initial state, **c**, the state when the piston is lowered, and, **d**, after the piston is retracted. **e**, Slice through a tomography measurement of the as-received carbon felt. Dashed box indicates the analyzed control volume. **f**, Slice through a tomography measurement of the as-received carbon foam. Dashed box indicates the analyzed control volume. **g**, Control volume shown in **e**, after piston has been lowered to 0.6 mm above the felt. **h**, Control volume shown in **f**, after the piston has been moved to 1 mm above the foam. **i**, Fraction of filled pores depending on the pore radius for felt and foam. **j**, Distribution of pore filling showing that for both materials most pores are either nearly completely filled or not filled at all.

## 2.3 Forced Wetting

Studies with specifically designed current collectors have shown that alkali-metal retention and release play an important role in efficient cycling of batteries with liquid alkali-metal negative electrodes [12]. Lightweight and cost-effective carbon materials could be an attractive conductive material in electrode compartments if they achieve a similarly high retention and release of the alkali metal. Here we want to learn how as-received carbon materials interact with liquid alkali metals under geometric constraints. We use a piston to push NaK to infiltrate into the open pore space of carbon foam and felt, in a custom-built XCT cell (**Figure 3a**). **Figure 3b-d** show schematic drawings detailing the pistons movement. Tomography slices are shown in pairs where the upper images (**Figure 3e-f**) each show the initial state. The dashed box indicates where the control volume is located that was used for quantitative analysis. It is deliberately chosen to represent the bulk volume of the porous carbon hosts. **Figure 3g-h** each show the control volume when the piston is lowered to 0.6 mm and 1 mm above the felt and foam, respectively, corresponding to the state sketched in **Figure 3c**.

It can be seen that the NaK partially wets the carbon host, filling some pores completely while not entering other pores at all. Significantly more NaK has entered the bulk volume in the foam as compared to the carbon felt. This is also confirmed when analyzing the fraction of filled pores, shown in **Figure 3i**. The carbon foam has a significantly greater fraction of filled pores with on average 39% of pores filled, while for the felt only 5.2% of pores are filled. **Figure 3j** shows that for both samples there is a clear divide between unfilled and mostly filled pores. As expected with the overall higher fraction of filled volume, the fraction of nearly completely filled pores in the 90% -100% filled decile is 40% for the foam and only 2% for the felt.

The significant difference in fraction of filled pores between both carbon samples correlates with their difference in pore size distribution. Larger pores enable easier NaK transport into the bulk volume of a porous carbon sample. This is in-line with classical wetting theory along the Young-Laplace equation, where the capillary pressure resisting the filling of a phobic pore is inversely proportional to the pore radius [26]. The results also indicate that in both cases the repellent nature of the carbon structure persists even when forced into contact with the NaK. This is shown through the limited pore filling even for the open porous carbon foam, but can also be seen from the preferential filling of different cavities such as between the carbon sample and the tomography cell's wall. In both samples, there is a moderate trend of larger pores being filled more often with NaK. This visual trend is confirmed by correlating the filled fraction of each pore with its pore radius. The covariance between pore radius and pore filling for the foam is considerably higher at 0.41 than for the felt at 0.044. It is worth noting that the pore geometry is rather different between both samples, with the carbon felt's pores generally not conforming the assumed spherical geometry on which basis the pore radius is calculated. This assumption is more applicable for the carbon foam sample. Pore geometry could also have an influence on the pore filling mechanism.

For both samples, when retracting the piston, the status of filled pores does not significantly change. This suggests that some energy barrier exists to complete dewetting and mechanical reversal of pore filling even for carbon materials phobic to NaK. It has been shown in other contexts of non-wetting fluids (in our case NaK) trapped in pores that a considerable pressure from a wetting fluid (in our case the gas) is needed to remove firmly trapped non-wetting fluid volumes [27]. Additionally, on structured surfaces, the impalement of liquid volumes by structural elements has been shown to enhance adhesion which needs additional energy input to be reversed, even for non-wetting liquids [28], [29].

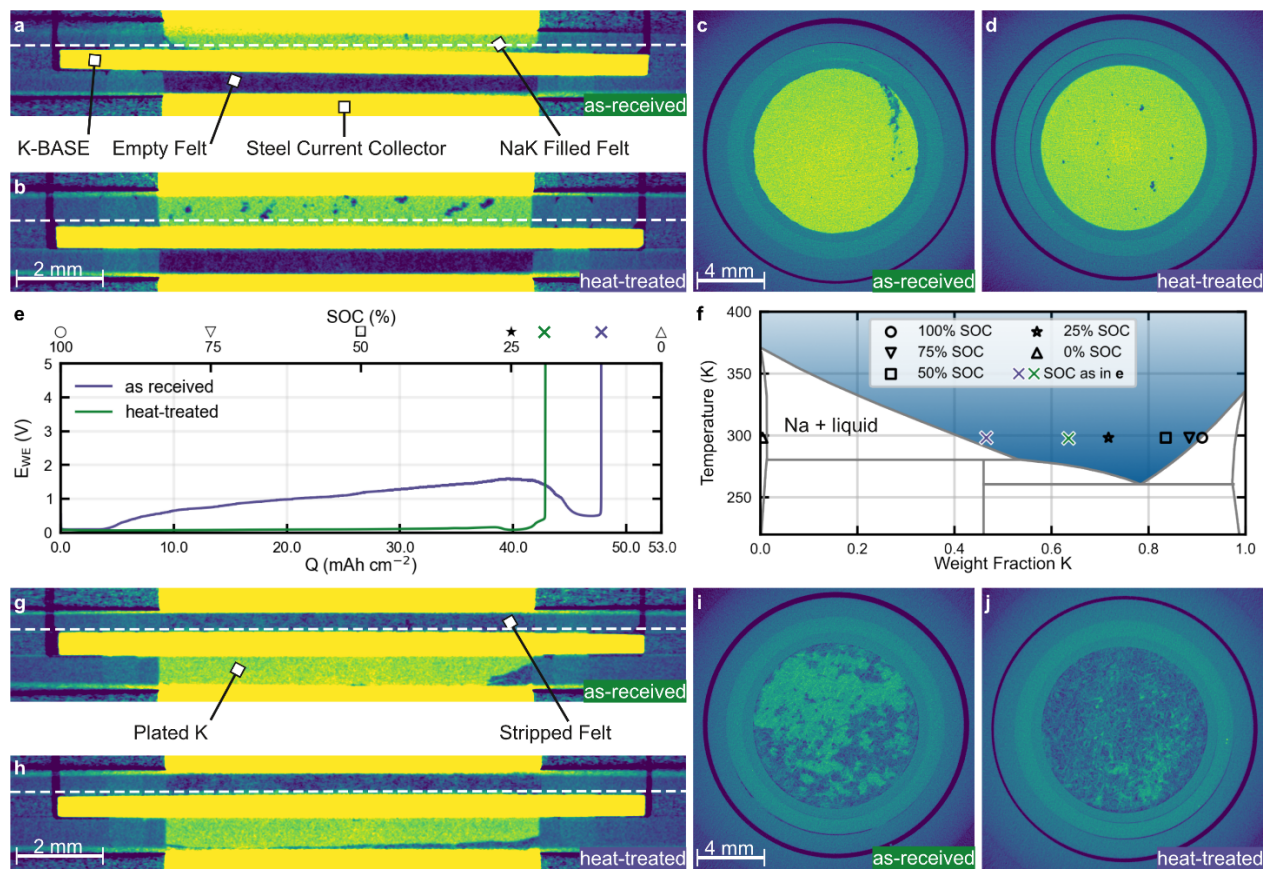

**Figure 4. Comparing electrochemical performance of as-received and heat-treated carbon felt. a**, Vertical slice through an XCT scan of an electrochemical cell with as-received carbon felt on both sides and an initial volume of Na10K90 in the upper compartment before discharge. The compartments are separated by a potassium-β''-alumina solid electrolyte (K-BASE). **b**, Vertical slice through an XCT scan of an electrochemical cell with heat-treated carbon felt on both sides and an initial volume of Na10K90 in the upper compartment before discharge. **c**, Horizontal slice through the same cell pictured in **a** on the height of the white dashed line. **d**, Horizontal slice through the same cell pictured in **b** on the height of the white dashed line. Substantial differences in alkali-metal distribution are visible between the heat-treated and as-received samples. **e**, Constant current discharge curves at 0.5 mA cm$^{-2}$ for electrochemical cells with as-received and heat-treated carbon felt samples. **f**, Sodium-potassium phase diagram with theoretical state of charge (SOC) values and their respective composition marked at room temperature. Additional purple and green crosses show the computed NaK composition at the end of the discharge curves shown in **e**. **g**, Vertical slice through an XCT scan of an electrochemical cell with as-received carbon felt after most of the initial potassium has been electrochemically transferred to the lower compartment. **h**, Vertical slice through an XCT scan of an electrochemical cell with heat-treated carbon felt after most of the initial potassium has been electrochemically transferred to the lower compartment. **i**, Horizontal slice through the same cell pictured in **g** on the height of the white dashed line. **j**, Horizontal slice through the same cell pictured in **h** on the height of the white dashed line. The position of the dashed line in all cells is 120 μm up from the K-BASE on the site where the initial volume of NaK is or was located.

## 2.4 Porous Carbon Hosts in Electrochemical Cells

In a next step we used differently treated carbon felt as conductive porous hosts to retain liquid NaK in the electrode compartments of an electrochemical cell. To this end, we assembled lab-scale solid-state battery cells with either as-received or heat-treated carbon felt on either side of a K-BASE disk. We added an initial volume of 80 µl of potassium-rich liquid NaK alloy (Na10K90) to the upper electrode compartment. These 80 µl NaK weigh around 67.5 mg for both cells. At the given composition this amounts to a theoretical areal charge density of 53 mAh cm$^{-2}$. For this experiment, we applied a stack pressure of approximately 1.5 MPa to the cell, which is significantly lower than in most solid-state battery research [3], [5], [30]. XCT scans of cells with this initial configuration are shown in **Figure 4a-d**, distinguishing as-received and heat-treated carbon felt. It can be seen that the initial distribution of liquid NaK in the upper cell half is different between as-received and heat-treated samples. While in the as-received carbon felt the gas remaining in the electrode compartment is one continuous volume, in the heat-treated carbon felt the spontaneous wetting leads to a uniform distribution of small gas bubbles throughout the entire volume of the felt.

**Figure 4e** shows a discharge experiment with cells assembled in that way with a moderate current density of 0.5 mA cm$^{-2}$ and a cutoff voltage set to 5 V. Both, as-received and heat-treated carbon felt are capable of providing an areal charge density of greater than 40 mAh cm$^{-2}$. In case of the heat-treated carbon felt, the achieved discharge capacity represents 80% of the available theoretical capacity. The mean overpotential before the onset of the steep final voltage increase is 88 mV. For the as-received carbon felt, the discharge capacity reaches 90% of the theoretical maximum, while the mean overpotential is more than ten times higher at 963 mV. This is most likely due to a larger contact area between the NaK and the SE for heat-treated carbon felt due to the more uniform NaK distribution.

During the discharge process, the liquid NaK alloy changes in composition due to potassium ions being transferred to the other electrode. This compositional change is highlighted in the phase diagram in **Figure 4f** with various symbols corresponding to different states of charge (SOC). These SOC symbols are also shown in the top axis of the diagram in **Figure 4e**. 100% SOC is calculated from the initial volume of Na10K90 in the electrochemical cells. The calculated composition at the end of both discharge curves is also indicated in the phase diagram. According to the phase diagram with the indicated initial NaK composition no solid Na should form up until lower than 10% SOC, corresponding to a composition of Na53K47. However, after disassembly, the alkali metal left on the SE surface on the upper compartment's side, unexpectedly seemed at least partially solid, demonstrating that the composition has changed enough to have entered the mixed phase-field of liquid NaK and solid Na. Possible reasons for this are twofold: Kinetic effects during discharge can lead to small volumes of liquid NaK becoming disconnected from the rest of the liquid alloy, preventing the compensation of removed potassium close to the SE, subsequently leading to the depletion of practically available potassium and as a consequence to the rise of overpotential [11], [31]. This is in line with the forced wetting experiments displayed in **Figure 3** showing practically no dewetting after retracting the piston. Another possibility is the removal of overall available potassium due to potassium intercalation into a graphitic portion of the carbon felt. Heat-treated and as-received felt might intercalate potassium at different rates and up to different amount, which could explain some of the difference in utilized potassium during the discharge. When most of the liquid NaK is depleted, this potassium would be hard to utilize electrochemically. These phenomena can shift the calculations regarding the composition of the still available NaK towards the left, sodium-rich side of the phase diagram as some potassium is practically removed from the NaK.

Tomography measurements taken after the discharges displayed in **Figure 4e** are shown in **Figure 4g-j**. For both cells, there is almost no alkali metal left in the upper compartment. The alkali metal plated into the bottom compartment is solid potassium and does not exhibit any obvious voids or pores. The stripped carbon felt in both cells is slightly higher absorbing, i.e., has a higher density as compared to carbon felt that has not been in contact with alkali metals in the bottom compartment of the freshly assembled cells in **Figure 4a-b**. This hints at the possibility mentioned above, of some potassium permanently reacting with the carbon felt, possibly forming graphite intercalation compounds. Such reactions have been reported in the literature for liquid potassium [32], showing that potassium-graphite intercalation compounds may form spontaneously.

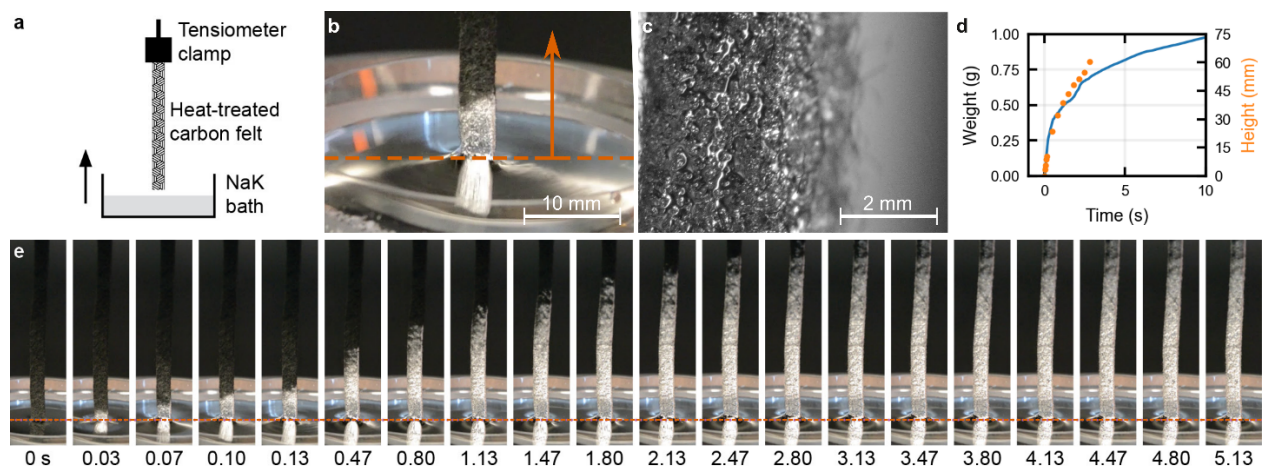

**Figure 5. Capillary action of heat-treated carbon felt. a**, Schematic of the experimental setup in a tensiometer, raising a beaker with Na23K77 to come in contact with a heat-treated strip of carbon felt, measuring the height to which the NaK is transported up the felt, and the weight gain. **b**, still image shortly after the felt touches the NaK showing a few millimeters of transport (see **Video S1**). **c**, Microscope image of the carbon felt soaked with NaK during the measurement (see **Video S2**). **d**, Cumulative weight (blue) and transported height (orange) of the NaK in the heat-treated carbon felt. **e**, Time series of still images documenting the capillary transport of the liquid NaK alloy through the felt.

### 2.5 Capillary Action of Heat-Treated Carbon Felt

To demonstrate the ability of heat-treated carbon felt to transport liquid alkali metal over a distance of 10 cm, an ex-situ experiment with carbon felt showed spontaneous wetting and subsequently quick pore filling. We brought a 10 cm long strip of heat-treated carbon felt into contact with a beaker filled with Na23K77 in a tensiometer setup. The general setup is sketched in **Figure 5a**. A video (**Video S1**) of the process is used to determine the speed of the capillary transport with the apparent surface of the liquid NaK defined as a height of 0 mm. The defined zero height line is indicated in **Figure 5b**. **Figure 5c** shows a still image from a closeup slow-motion video of the felt being filled (**Video S2**). Even after the initial capillary transport front has passed the field of view of the microscope camera, the video still shows progressive infiltration of the remaining pore space. Data showing the transported height visually measured from the video footage and weight uptake measured by the tensiometer is displayed in **Figure 5d**. **Figure 5e** shows the corresponding timeseries of different still images taken from the video. It can be seen that after around 5 s the NaK reaches the top end of the carbon felt. The weight indicated in **Figure 5d** keeps increasing for a longer duration with more and more pores being filled over time.

## 3. Conclusion

Porous carbon hosts can be tuned to drastically change their wetting interaction with liquid NaK alloy. For the stable operation of liquid alkali-metal batteries, the electrical contact throughout the negative electrode compartment and an intimate contact between the liquid metal and the SE need to be maintained at all times. To help achieve these goals, porous carbon materials can offer a three-dimensional conductive framework combined with effective transport of liquid alkali metals through capillary action. We showed that the effect of heat-treatment of carbon materials applies to samples of very different pore size distributions. We used XCT to gain insight into the volume distribution and to quantify alkali-metal wetting of porous carbon in 3D within a cell. We found that spontaneous wetting in porous carbon hosts displaces most of the gas volume with NaK showing a moderate dependence on the pore size. Pore filling in forced wetting conditions is not reversible for the tested carbon hosts while showing a large dependence on the carbon hosts morphology. Initial results in electrochemical cells show that electrodes made up of NaK stored in porous carbon hosts can provide a large fraction of the theoretical capacity even at very high areal capacities of 40 mAh cm$^{-2}$ and low stack pressures. For these experiments, using a heat-treated carbon felt could reduce the resulting overpotential by an order of magnitude. Finally, heat-treated carbon felt has also shown the capability of transporting liquid alkali metal over multiple centimeters in seconds. In summary, low-cost and flexible carbon materials can provide key advantages when used as structural elements in liquid alkali-metal negative electrodes.

## 4. Materials and Methods

*Materials:* Commercially available carbon materials were used: The carbon felt was sourced from Alfa Aesar (3.18 mm thickness, 99.0% purity), while the carbon foam stemmed from ERG Aerospace Corp. (reticulated vitreous carbon, 100 ppi, 3% relative density). The alkali metals used were purchased from thermo scientific: Potassium 98%, chunks in mineral oil, and sodium 99.8% oiled sticks, wrapped in aluminum foil, respectively.

*Carbon host preparation:* The samples were cut out to the diameter of the respective tomography or electrochemical cell (4 mm, 6 mm, or 10 mm), from larger pieces of carbon foam or felt using a hole punch. For the capillary action test, a 10 cm strip of carbon felt was cut from a larger piece. The thermal activation process was conducted in a muffle oven in air at 500 °C for 3 hours, with a heating rate of 5 °C/min and natural cooling, transferring the samples into the inert gas atmosphere of a glovebox before they were fully cooled. [17]

*Alloy preparation:* Eutectic NaK ($Na_{0.32}K_{0.68}$) and potassium rich NaK alloy ($Na_{0.16}K_{0.84}$) were prepared by combing metallic sodium and potassium in a glass vial by weight [33]. Due to the low melting point of eutectic NaK alloy, K and Na melt upon contact. A fully liquid alloy was achieved by manual mixing in the glass vial. With the naming scheme based on the weight percentage of NaK alloys, the eutectic alloy is referred to as Na23K77 and the potassium rich alloy as Na10K90.

*Cell preparation:* Custom-made tomography cells were assembled in a glovebox under argon atmosphere. First, the porous carbon host was added. Next, the liquid alkali metal was added on top of the carbon host using a pipette. The cells were sealed by commercially available rubber gaskets (EPDM or FKM) to prevent oxidation of the alkali metal once outside of the glovebox. For the electrochemical cells, a potassium β''-alumina solid electrolyte (K-BASE) was inserted into the cell first, securing it with a screwed inset, insulating both electrodes from each other with EPDM rubber gaskets. A porous carbon host was inserted into the bottom of the cell and the stainless-steel current collector was inserted and pushed to gain contact with the carbon. For the side initially containing liquid NaK alloy, 80 µl of the alloy was pipetted in first, subsequently adding the carbon host and current collector. Airtight sealing to the environment was achieved by a PTFE compression fitting between the current collector and the cell's PEEK body. Finally, the electrochemical cell was placed in an aluminum frame, setting the stack pressure of approx. 1.5 MPa by tightening a screw that applies a uniaxial force perpendicular to the K-BASE to 0.2 Nm torque. The relationship between torque and stack pressure was verified by using a pressure gauge.

*Electrochemical testing*: Electrochemical discharge experiments were conducted in an argon filled glovebox, using a BioLogic VSP-3 potentiostat. After an initial potentiostatic electrochemical impedance spectroscopy (PEIS) a constant current step with a current density of 0.5 mA $cm^{-2}$ and a cutoff potential of 5 V was carried out. After a successful discharge step, another PEIS measurement was conducted to confirm no short circuit had formed in the cell (**Figure S3**).

*X-ray tomography:* Tomography experiments to evaluate forced and spontaneous wetting were conducted at the tomography end station of the high energy materials science (HEMS) beamline P07 at the synchrotron radiation facility PETRA III at Deutsches Elektronen-Synchrotron DESY,

Hamburg, Germany. The energy of the X-ray beam was 81 keV monochromatized using a double Laue monochromator. X-rays were detected by an indirect detector system using a 100 µm thick CdWO4-scintillator and a camera equipped with a CMOS sensor that was developed within a cooperation by Hereon and Karlsruhe Institute of Technology (KIT) [34]. Two slightly different parameter settings were used in the following. The scintillator was placed at a distance of 300 mm and 150 mm, respectively, from the sample. The raw images have 5120 × 1751 pixels and 5120 x 1540 pixels, respectively, with an effective pixel size of 1.27 µm due to the optical system with fivefold magnification. 5001 and 6000, respectively, projections were obtained during a 180° fly-scan with an exposure time of 100 ms and 200 ms, respectively. Images were binned twofold during reconstruction. Filtered backprojection with a Ram-Lak filter was applied for tomographic reconstruction following the workflow of the reconstruction and analysis tools for tomographic data implemented in MATLAB [35], [36]. The ASTRA toolbox was used for tomographic backprojection [37], [38].

Tomography measurements of the electrochemical cells were conducted with an RX Solutions Easytom lab scale tomography device (cone beam setup) with an accelerating voltage of 130 kV, beam current of 76 µA and a PaxScan 2520DX-I flat panel detector with a pixel array of 1,536 x 1,920 pixels was used. The pixel size was set to be 11 µm, 1856 projections have been captured over 360° with an exposure time of 1 second for each projection. For the drift correction, filtering and reconstruction the software package Xact (RX Solutions) was used. The underlying reconstruction is based on the filtered back projection algorithm.

*Tomography data handling:* Reconstructed and binned tomography data was further prepared for segmentation through custom made python scripts utilizing commonly used scientific python packages [39], [40], [41]. The preparation included further binning and cropping, artifact removal through filtering, and registration of multiple measurements of the same sample, if a sample was too large to fit in the 1.8 mm imaging height in z-direction. Segmentation was performed in ilastik, using its pixel classification workflow and underlying random forest classifier [24]. Finally, from the segmented volumes, pore identification was performed utilizing MorphoLibJ's distance transform watershed function with further analysis of the resulting data again being done in custom made python scripts utilizing the same python packages as mentioned above.

*Tensiometry setup*: The tensiometer device was a dataphysics DCAT25 set up in an argon filled glovebox. To accommodate the long carbon felt sample, a custom-made clamp was used to attach the sample to the integrated scale, which records 50 datapoints per second. A Nikon D3100 digital camera was used to capture the video footage from outside of the glovebox.

## Acknowledgments

This study has received funding from the Bundesministerium für Forschung, Technologie und Raumfahrt under the grant No. 03XP0516. Additional funds were received from Fonds der Chemischen Industrie im Verband der Chemischen Industrie e. V. under the grant No. 661737. This research was supported by the Federal Federal Ministry of Research, Technology and Space (BMFTR/BMBF) as part of the FORBATT project "TomoFestBattLab", grant number 03XP0462, and by the BMFTR/BMBF project "NATTER" grant number 03XP0525C. We acknowledge Deutsches Elektronen-Synchrotron DESY (Hamburg, Germany), a member of the Helmholtz Association HGF, for the provision of experimental facilities. Parts of this research were carried out at PETRA III. Beamtime was allocated for proposals I-20230831 and I-20240475. This research was supported in part through the Maxwell computational resources operated at DESY. We want to thank Lars Drescher and his team for his support with cell design and manufacturing.

## Data availability

The data and code that support the findings of this study are available from the authors upon reasonable request.

## Conflict of Interest

The authors declare that they have no conflict of interest.

ToC Content

Porous carbon host materials are investigated for their possible usage in liquid alkali-metal negative electrodes. X-ray computed tomography reveals pore-filling behavior under various conditions: non-wetting, forced wetting, and spontaneous wetting. The applicability of porous carbon in electrochemical cells is demonstrated at high areal capacities exceeding 40 mAh cm$^{-2}$. Additionally, capillary-driven transport over distances of several centimeters is shown.

**Wetting Interactions Between Porous Carbon Hosts and Liquid Sodium-Potassium Alloys Toward Their Use in Negative Electrodes of Alkali-Metal Batteries**

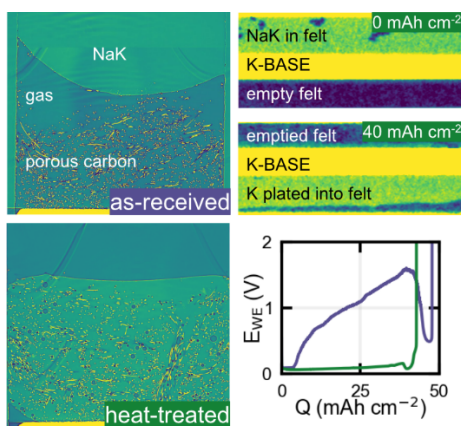